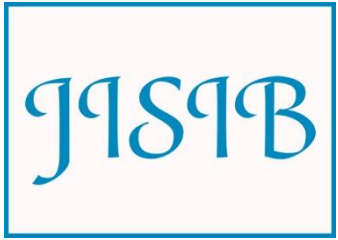

# Patents used by NPE as an Open Information System in Web 2.0 – Two mini case studies

Abdelkader BAAZIZ[1], Luc QUONIAM[2]

[1]Aix-Marseille University, France

[2] University of Sud Toulon Var, France

Email: kbaaziz@gmail.com, mail@quoniam.info

**ABSTRACT:** The Information Systems around patents are complex, their study coupled with a creative vision of "out of the box", overcomes the strict basic functions of the patent. We have, on several occasions, guiding research around the patent solely-based on information, since the writing of new patents; invalidation of existing patents, the creation of value-added information and their links to other Information Systems. The traditional R&D based on heavy investments is one type of technology transfer. But, patent information is also, another powerful tool of technology transfer, innovation and creativity. Indeed, conduct research on the patent, from an academic viewpoint, although not always focusing only on financial revenue, can be considered as a form of "Non Practicing Entities" (NPE) activity, called rightly or wrongly "Patent Trolls".  We'll see why the term "patent troll" for this activity is controversial and inappropriate. The research we will describe in this paper falls within this context. We show two case studies of efficient use of patent information in Emerging countries, the first concern the pharmaceutical industry in Brazil and the second, the oil industry in Algeria.

## 1.0 Introduction

*"Whoever finds what he seeks, he has done generally a good job as a schoolboy, focusing on what he wants, he often neglects the signs, sometimes small, that bring something over than the object of his forecasts. The true researcher must pay attention to signs that reveal the existence of phenomenon that he does not expect."* [1]

This quote (freely translated) of the French physicist Louis Leprince-Ringuet (1957) shows perfectly the research that can be conducted on open data and particularly on "patent" information.

The Information Systems around the patents are complex, their study is coupled with a creative vision of "out of the box" (Swinner & Briet, 2004), overcomes the strict basic functions of the patent. We have, on several occasions, guiding research around the patent since the writing of new patents, invalidation of existing patents, the creation of value-added information and their links to other Information Systems (Quoniam, 2013).

The patent is undeniably one of the more important tools of technology transfer, innovation and creativity. Indeed, talking about patent is talking about Research & Development (R&D). However R&D without marketing is expensive. Marketing is also expensive, where the needed steps of "licensing" are worth to dispense with R&D or marketing. This is one kind of technology transfer. But "generic", "public domain", "open data" and "reverse engineering" are other forms of technology transfer, without necessarily financial reward. Thinking that way is looking at the patent with other eyes (Quoniam, 2013). Those who act in this manner are called rightly or wrongly, "Patent Trolls" or, in less pejorative term "Non Practicing Entities (NPE)".

## 2.0 Patent Trolls & Non Practicing Entities

A Non Practicing Entity (NPE) is a person or company that amasses patent rights. The patents typically belong to a single technological field, or a grouping of related technologies. The NPE does not practice the patents, meaning that the NPE does not produce any goods or provide any services based on the patents rights that are held (Halt & al., 2014).

A patent troll is defined as one type of NPE. Patent trolls use the licensing and patent litigation as a business model (Quoniam, 2013). They purchase large numbers of patents, often from bankrupt firms, with the intention of launching patent infringement suits against companies and individuals that they maintain have illegally used some element of something for which they hold the patent. A highly publicized case was that of Research in Motion (RIM), manufacturer of Blackberry mobile phones, which was ordered to pay $ 612.5 million to New Technology Products (NTP) to stop the litigation instigated to the Courts.

However, patent trolling is not a new phenomenon. Already in 1878, Senator Issac Christiancy seemed to have patent trolling in mind. He rightly noted:

*"Among a host of dormant patents, some will be found which contain some new principle ... which the inventor, however, had failed to render of any use in his own invention. And some other inventor, ignorant that such a principle had been discovered . . . had the genius to render it of great practical value ... The patent-sharks among the legal profession, always on the watch for such cases, go to the first patentee and, for a song, procure an assignment of his useless patent, and at once proceed to levy blackmail upon the inventor of the valuable patent."* [2]

In fact, the term "Patent Troll" appeared in the late 1990s and was used at least once in 1993 with a different meaning, to describe countries that file aggressive patent lawsuits.

Excessive patent protection by the big firms may hamper further innovation, especially when they limit access to essential knowledge, as in the case of emerging technological fields. In this context, too broad a protection on basic inventions can discourage follow-on inventors if the holder of a patent for an essential technology refuses access to others under reasonable conditions. This concern was often expressed for new technologies in the fields of genetics and software (OECD, 2004).

---

[1] Louis LEPRINCE-RINGET in "Des atomes et des homes", Fayard, Paris, 1957, Page 57.

[2] As quoted in Gerard N. Magliocca, Blackberries and Barnyards: Patent Trolls and the Perils of Innovation (2007), Notre Dame Law Review, June 2007. Available online: http://ssrn.com/abstract=921252.

### 3.0 "The Good, the Bad and the Ugly"

Conduct research on patents, from an academic viewpoint, although not always focusing on financial revenue, can be considered as a form of NPE activity (Quoniam, 2013). The research that we will describe in this paper falls within this context.

We'll see why the term "patent troll" for this activity is controversial and inappropriate:

- First, this activity may be described as "Soft Research and Development" because it relates to innovation i.e., conducting R&D, discovering and creating new knowledge without the traditional R&D laboratories, but solely from the information available in open data sources including the patent databases;
- Second, from research based solely on the information, it can lead us to invalidate existing patents or write new patents in a given technical field;
- Third, this research could be described as Research Social Responsibility (RSR), with legal and legalistic action, similar to Corporate Social Responsibility (CSR) activities for academic research;
- Fourth, it refers to an unconventional form of thinking "out of the box" by setting strong links between "hard technologies" and "soft technologies" (Jin, 2005), and establishing transitions from one to the other and vice versa, based on open data sources and patent information.

Shrek, a good friendly ogre and famous in the movies industry, has supplanted all the bad ogres and other ugly trolls of medieval legends. This caricatured picture illustrates the differences between the forcing exerted by the bad ogres (the big firms that create barriers to innovation), the blackmail used by ugly patent trolls in the common sense of the term (licensing with purely financial goal), and actions taken by the good patent trolls (technology transfer mediator, knowledge disseminator and know-how sharer). We therefore suggest a new term for naming the good trolls, why not Patent Robin-Hoods?

### 4.0 Conceptual frameworks and foundations

Intellectual property consists of two parts: copyright and industrial property rights. The latter includes inventions (patents), trademarks, industrial Designs and indications of geographical origin.

The relationship between intellectual property and economic development is obvious and has been the subject of many publications. It is part of the tangible manifestations of intangible activities related to "knowledge societies" (Binde, 2005).

In Industrial Property, Patent plays a key role, by its strategic importance as it represents a property right for an invention, for a product or process that provides a new technical solution to solve a problem. The conditions for obtaining such ownership, conditions of validity of these rights and how to enforce them, are described in the literature (Quoniam, 2013).

We are interested here in a patent from a strictly informational viewpoint and opportunities for exploitation thereof for purposes other than strict property rights. The patent is seen as a way to communicate to the market purely technical and technological research. We show that the patent can be a way to conduct research, well beyond strict technical research. It allows you to work on many fields related to "soft technologies" (Jin, 2005) and put into perspective the evolution of "hard technologies" to "soft technologies".

### 5.0 Criticism of Maslow's model (1943) & Aziz Ungku's model (1983)

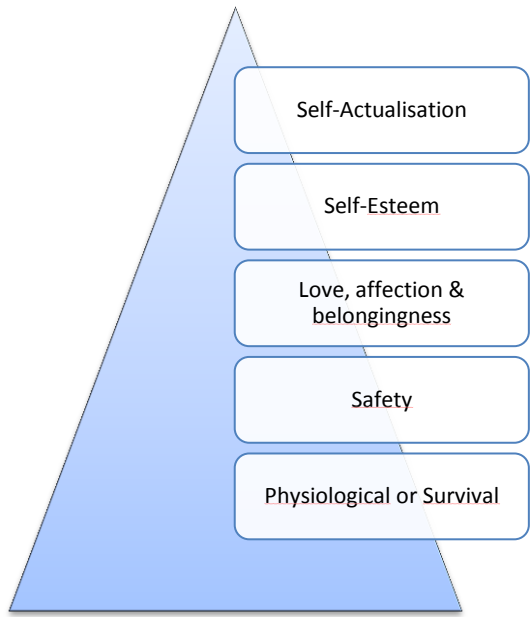

Fig. 1. Maslow's pyramid (1943)

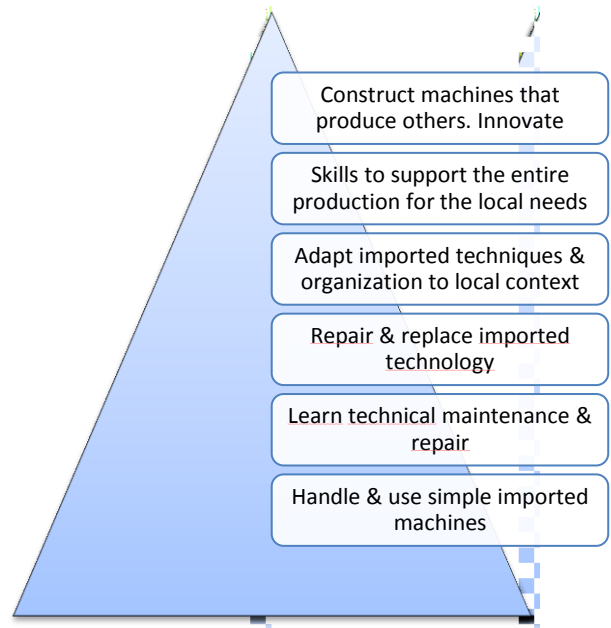

Fig. 2. Ungku Aziz's Pyramid (1983)

The main criticism of Maslow's model is based on the assumption that the individual pass from one level to another only once he has satisfied the needs of lower level, yet every human being does not necessarily have this mode of prioritization of needs, not in his personal or professional life. Observations in organization show inversion or co-existence of levels. Also, Maslow's model cannot explain the lack of motivation. Thus, according to this model, it's not possible to do a good job with a small salary. Conversely, an excellent salary does not guarantee a strong motivation.

In our humble opinion, the same main criticism is opposable to the Aziz's model. This means that the organization pass from one stage to another only once satisfied with the requirements of a lower stage. Each of the stages of technological evolution considered as a prerequisite to the successful realization of the preceding stage. Thus, according to this model, there is no way of jumping straight into the higher stages and bypassing the earlier preparatory stages (Idris, 2000).

Using patent information integrated in Information Systems field, we demonstrate the invalidity of these models.

In "The Creative Evolution", the French philosopher Henri Bergson (1907) defines the concept of "homo faber" as (freely translated):

*"... Intelligence, in its original sense, is the faculty of creating artificial objects, especially*

*tools to make tools, and to vary indefinitely its makings."* (3)

This approach applies to patents and was adopted by the World Intellectual Property Office (WIPO) which accurately describes the ultimate stage of technological development as defined by Ungku Aziz (Idris, 2000; Idris, 2003):

*"The sixth stage consists in learning to make machines that produce machines, as well as learning to innovate and being ready to approach the frontiers of modern technology in such fields as computers, robotics and biotechnology, using energy and raw materials without causing irreparable damage to the environment, and becoming an exporter of high technology products. The intellectual property system is already integrated into R&D activities."* (4)

---

3 Henri BERGSON in "L'évolution créatrice", p. 100, originally published in 1907. Les Presses Universitaires de France, 1959, Available online: http://classiques.uqac.ca/classiques/bergson_henri/evolution_creatrice/evolution_creatrice.html
4 Ungku AZIZ, "Must patterns of change in developing countries follow the West? What other possible patterns?" In technological innovation: Universities of the commonwealth, Birmingham (August 1983). Cited by Kamil Idris (WIPO) in "A Brochure on Intellectual Property Rights for "Tool to make tools transform its environment, shapes it with his feeling and his hands". This brings us closer to the concept of "stigmergy" [5], addressed by Charles Victor Boutet for the Information Sciences in light of the web 2.0 (Boutet, 2011), and defined as a form of self-organization that produces complex, seemingly intelligent structures, without need for any planning, control, or even direct communication between the agents (Wikipedia).

To use the patent as an object of research, we need to mobilize these concepts to overcome the available tools for the patent analysis and create other tools, to analyze machines that create tools and increase the degree of complexity.

The European Patent Office proposes its base (over 70 million patents) and an Application Programming Interface (API), a tool which has already created many tools but offer to Researchers in Information Systems the ability to create tools to create tools that help understand and transform the environment. The only limit to analysis with the API is our own imagination.

It is therefore not necessary to follow all the stages recommended by Ungku Aziz to reach the ultimate stage. For us, a research activity is situated around the concept of "homo faber", so as not to confine research activities to the role of passive spectator, without interaction with the society that houses them. It also responds to the need to conduct research into multi-skilled teams. In the field related to patents, it quickly becomes indispensable.

We will show along this paper, the contribution of Science Information, but also knowledge on law of Industrial Property, analyzed technical field (chemistry, materials, pharmaceuticals, medicine), Information Technology, Corporate Social Responsibility, Sustainable Development, Innovation, creativity, without being exhaustive, are quickly needed, forcing them to work in multi-skilled teams, which often justifies multi-author approaches in the literature.

## 6.0 Patent Information as Open Information System

The International Patent Classification is a unique system to show technology. Each patent must be described in it. As a corollary everything "patentable" is described by it. It was set-in place by the Strasbourg Agreement concerning the International Patent Classification of March 24, 1971, amended September 28, 1979.

IPC was developed in two "official languages: French and English", divides technology into eight sections with approximately 70,000 subdivisions described by symbols. This classification is useful to search on patent documents in the context of research on the "State of Art".

A patent database, free and freely accessible online, allows to do research in this classification. Thesaurus allows browsing in the classification with 22798 English keywords and 25676 French keywords.

IPC is the core of an interface between different languages for describing "patentable subject matter". It is the means to see what is patented around a subject matter in another language without knowing the language. It can also enable quick dialogue between experts communicating in different languages, in the same field.

Generalizing a little more, it is technically possible to describe an entire problematic. This is the case of the Green Inventory according to IPC, developed by the Experts Committee of IPC Union based on Climate Change Mitigation Technologies (CCMTs) in order to facilitate the search for information on patents related to Environment Technology. This green inventory was built with a list of terms established by the United Nations Framework Convention on Climate Change (UNFCCC). This application is proof of using IPC as a pivot system between Societal and Environmental issues and their declination to industrial realities, providing tracks of application into the market realities of societal evolution as "Sustainable Development" or "Corporate Social Responsibility". This is also a crossing point between "hard technologies" and "soft technologies".

---

Universities and R&D Institutions in African Countries", June 2000, ISBN 92-805-1097-7

[5] The term "stigmergy" was introduced by French biologist Pierre-Paul Grassé in 1959 to refer to termite behavior. He defined it as stimulation of workers by the performance they have achieved, and captures the notion that an agent's actions leave signs in the environment, signs that it and other agent's sense and that determine and incite their subsequent actions (Wikipedia).

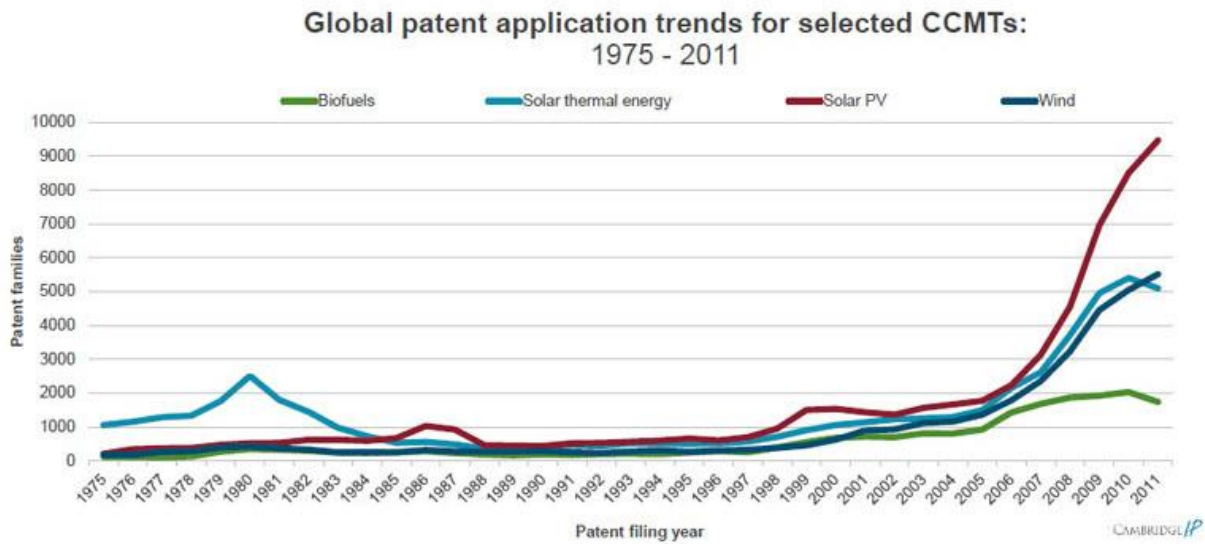

Fig. 3. Patenting activity is taking off in several green-tech sectors. (Source: WIPO, 2014).

This example is adaptable to many other sectors. It could be an investigation framework of reconciliation between public and private sectors, in order to search funding under development based on the "triple helix" model (Etzkowitz, 2008) where synergies "government - research - private sector" are sought.

During his study of the bibliography of a theses sample (586 from Brazil and 580 from U.S.) in the field related to technology research, Juliana Ravaschio found only 15% of theses that have patent citations (Ravaschio, 2010). This result shows the lack on "patent" knowledge in the academic world. Lack in part, due to the use of different vocabularies and concepts, cleavage of communication. It was on this ground that the research "Automatic mapping of scientific and technical bibliographies databanks using the International Patent Classification (IPC): contribution to the rapprochement between sciences and technology" was developed by Pascal Faucompré. This research considers the IPC as a pivot system with keywords for indexing scientific databases, has set glances descriptions of academic research with an industrial description of these research (Faucompré, 1997).

A patent granted is legally valid at a given period. Beyond this period, the patent becomes invalid. Wanise Barroso (2003), a patent examiner in the Patent Office in Brazil upon completion of her research, worked on the subject. She proposed the creation of a sub-base of this patent office, containing only "public domain patents" for distribution to SMEs in order to facilitate technology transfer. This was possible by querying the database on the field which informed the legal status of the patent. She showed that more than 30% of the Brazilian Patent Base became in public domain in 5-6 years and around 40% of the base was in public domain which may be used freely as technological support, since applications patent must be written to be able to reproduce what is claimed. If we assume on equal proportions in other available databases, would bring to 28 million, the number of available patents in full text and freely "reusable" in the world as "Technical and Technological Encyclopedia".

This number does not include the geography to the territory for which it was filed and paid patent rights. Although having no precise estimate, the number becomes really significant. This fact can be used systematically to try for example to develop natural products or to add value to them. There is a source of development assistance, regardless of the country. At least it represents a source of inspiration and creativity almost without limit. Indeed, the free use of a patent is legal from the moment it falls into the public domain or is invalidated by the courts.

Technology disclosed in a patent document may be in the public domain if:

- The patent application has not been filed in a given country;
- The patent has not been granted;
- The patent term has expired, or the patent has not been renewed;
- The disclosed information is not covered by the claims.

In any case, the information is always in the public domain. According to Karnik (2008), the strategies for the invalidation of a patent which are followed by attorneys in the judicial courts are:

- Invention claimed in the patent is not novel;
- Subject of the claim of the patent is not an invention;
- Patent was wrongfully obtained by a person other than the person entitled;
- Insufficient disclosure of the invention;
- Obviousness;
- The claims included in the patent are not fully substantiated by the description provided;
- Failure to disclose information relating to foreign applications;
- Principle of "First to file – First to invent";
- Patent holder did not exercise diligence in pursuing the patent application process (Patent grace period).

### 7.0 Patent Information as Knowledge Database Discovery

To respond to this problematic: "how to find substances that interact with all the symptoms of a disease without anyone had made the relationship between these substances and disease?" Jean Dominique Pierret (Pierret, 2006; Pierret & al., 2010) conducted research on "methodology and structure of a knowledge discovery tool based on the biomedical literature". This research led to filing four new patents based on new medical indications of molecules previously patented in order to reach the market.

This proves that it is even possible to "invent" and "patent" by a recombination of existing knowledge on technical documentary barely developed. This kind of methodology could be generalized to materials, processes, etc.

The experience of Jean Dominique Pierret is to our knowledge, the first creation of "patentable subject matter" starting from documentary research. To go further in this field it will be necessary again to use the concept of "homo faber" to improve document interfaces to systematize this kind of queries.

The methodology called "Diseases Physiopathology Molecules (DPM)" is part of methods grouped under the term "Knowledge Database Discovery (KDD)". This is indeed, weak signals which Pierret (2006) attempted to show through the KDD process. The results are spectacular as they allowed Jean-Dominique Pierret and his colleagues of GALDERMA R&D, to file five (05) new patents based only on information:

| Titles | Publication numbers | Publication date |
|---|---|---|
| Use of a DIPYRIDYL Compound for treating ROSACEA | US20120322829-A1, CA2782048-A1, EP2506851-A1, WO2011064508-A1 | 20/12/2012 |
| Administration of TROPISETRON for treating inflammatory skin diseases/disorders | US2009048289-A1, CA2644458-A1, EP1993542-A1, WO2007099069-A1 | 19/02/2009 |
| Use of AZASETRON for the treatment of ROSACEA, and pharmaceutical compositions | WO2007138234-A1 | 06/12/2007 |
| Use of ZATOSETRON for the treatment of ROSACEA, and pharmaceutical compositions | WO2007138233-A3, WO2007138233-A2 | 06/12/2007 |
| Use of GRANISETRON for the treatment of sub-types of ROSACEA, and pharmaceutical compositions | WO2007138232-A3, WO2007138232-A | 06/12/2007 |

Table 1. Patents list filed by GALDERMA R&D based on Pierret's works

## 8.0 Case studies: Proof by doing

*"I hear and I forget. I see and I remember. I do and I understand."*
*Confucius, Chinese philosopher (551 BC - 479 BC)*

### 8.1 The Tenofovir case study (Health / Medicine fields)

The Federal Constitution of Brazil establishes, in Article 196, that health is a right of all and that it is the duty of the state to do this right to be guaranteed. This constitutional right is regulated by a 1990 law that, among other regulates the Unified Health System (UHS) to ensure full therapeutic assistance, including pharmaceutical assistance (Miguel & al., 2010). The cost of this legal provisions exceed R $ 1.4 billion (580 million €), with a number of pharmaceutical products distributed evolving from 15 in 1993 to 243 in 2007 (Carias & al., 2011).

Brazilian scientists were early interested in this (2003), using Information Systems to find ways in order to contain costs (Barroso & al., 2004) by a joint team of Brazilian and French researchers. They were interested in the Tenofovir case and they filed an opposition on patent application in conjunction with a research team in India for the antiretroviral used in the treatment of AIDS (Barroso & al., 2010).

The patent was invalidated and the implications of the research disclosed in the press[6]. It was important to manufacture this drug in a generic form, at significantly lower costs (50%), so more patients treated by the free health system and more released resources (R$ 110 million saving in 4

years) could be allocated for other types of free health access.

"For the first time, AIDS patients in developing countries will have access to the same drugs as those living in rich countries", says Philippe Douste-Blazy, UNITAID Director. This research, conducted without any funding since the financing was refused in France by the National Agency for Research on AIDS and Viral Hepatitis (believing that the research does not lead), was awarded an innovation prize in Brazil (Quoniam, 2010).

### 8.2 Drill-bit design case study (Mechanical Engineering / Oil & Gas fields)

This work conducted on behalf of an Algerian state-owned firm specialized in drilling tools manufacturing, led to a publication that shows the innovation opportunities offered by the use of reverse engineering combined with patent information in the oil & gas industry. A case study of drill bits design and optimization for oilfield drilling was proposed with outlining a parallel cognitive process associated to the technical process of reverse engineering (Baaziz & al., 2014).

The results of this work is a significant performance achieved during drilling operation of 12 ¼ section of BRNP#1, an exploration oil well located in Berkine east basin. On May 2014, the PDC tool 12 ¼ diameter, designed by the Engineering Team according to the process described above, has achieved this performance by drilling 769 meters in less than 34 hours and the rate of penetration (ROP) has approached the threshold of 23 meters per hour and peaked at 22,92 m/h without additions (connection). This is the best performance to date for such tools in Berkine basin area. The tool wear parameters were acceptable given the performance achieved which allowed to the client to reduce non-productive time (NPT) and costs of drilling operations. This study had multiple benefits for the firm and its client and, following the performance achieved:

- The used PDC tool is in its third remodelling (repair), which represents a saving of 40% of the price of a new drill bit;
- Cost savings for the client are of the order of USD 137,000.00 for 12¼ phase;
- Time recovery is over three days of drilling.

This study identified 7259 patents on the matter "drill bit" for the period from 1907 to 2013

---

[6] Various press:
- "Blanver entrega ao governo os primeiros lotes de genérico contra AIDS" in "Tribuna da Bahia Online", Available online: http://www.tribunadabahia.com.br/news.php?idAtual=81752 , May 2011.
- "Sida: le prix des médicaments baisse dans les pays pauvres" in « Le Figaro.fr », Available online: http://sante.lefigaro.fr/actualite/2011/07/12/11013-sida-prix-medicaments-baisse-dans-pays-pauvres , July 2011.
- "Funed produzirá genérico contra aids" in "O Reporter", Available online: http://www.oreporter.com/detalhes.php?id=40202 , February 2011.

including 2442 patents fallen into the public domain due to expiration of the protection period.

Baaziz & al. (2014) have rightly noted that it is interesting to verify the legal status of the patent legacy of firms currently being acquired or merged. This information can be verified by looking INPADOC legal status of patents. Indeed, the successive mergers and acquisitions incurred by the firms may generate dysfunctions in the intangible assets management. Patents can fall into the public domain due to these legal flaws.

### 9.0 Conclusion

This paper gives two examples from the field of Information Systems. It is for readers who would be interested to evoke new insights in patent matter and reciprocally, who in the field of Industrial Property would like to gain a deeper understanding of what could bring him the prism of Information Systems. It is also a plea for those who believe that there are alternative voices for technology transfer and innovation in both developed and developing countries through the use of patent information.

We've tried to demonstrate the usability of patent matter as a field of Research in Information Systems, by restricting around research experience and research directions that we may have in this field. It is obvious that given the vastness of the subject and its interconnections, it is not treated exhaustively here.

The integration of Patent studies in Information Systems field had the potential to support a development strategy that burns stages and borrows shortcuts in order to avoid heavy investments like "Hard R&D" processes. This contributes to reduce the gap between developed countries and developing countries.

Looking to the future, it seems that the possibilities of burning stages are even more promising through the opportunities offered by Web 2.0 technologies, which facilitate the flow of information. Many constraints of time and distance are abolished due to the variety of "open" formats used to disseminate knowledge and create links between Researchers and Professionals.